# Raman-gain induced loss-compensation in whispering-gallery-microresonators and single-nanoparticle detection with whispering-gallery Raman-microlasers

Şahin Kaya Özdemir[1*], Jiangang Zhu[1], Xu Yang[1,2], Bo Peng[1], Huzeyfe Yilmaz[1], Lina He[1]

Faraz Monifi[1], Gui Lu Long[2], Lan Yang[1*]

**Affiliations:**

[1] Department of Electrical and Systems Engineering, Washington University, St. Louis, MO 63130, USA

[2] State Key Laboratory of Low-dimensional Quantum Physics and Department of Physics, Tsinghua University, Beijing 100084, P.R. China

* Correspondence and requests for materials should be addressed to S.K.O (email: ozdemir@ese.wustl.edu) or L.Y. (email: yang@ese.wustl.edu)

**Abstract**. Recently optical whispering-gallery-mode resonators (WGMRs) have emerged as promising platforms to achieve label-free detection of nanoscale objects and to reach single molecule sensitivity. The ultimate detection performance of WGMRs are limited by energy dissipation in the material they are fabricated from. Up to date, to improve detection limit, either rare-earth ions are doped into the WGMR to compensate losses or plasmonic resonances are exploited for their superior field confinement. Here, we demonstrate, for the first time, enhanced detection of single-nanoparticle induced mode-splitting in a silica WGMR via Raman-gain assisted loss-compensation and WGM Raman lasing. Notably, we detected and counted individual dielectric nanoparticles down to a record low radius of 10 nm by monitoring a beatnote signal generated when split Raman lasing lines are heterodyne-mixed at a photodetector. This dopant-free scheme retains the inherited biocompatibility of silica, and could find widespread use for sensing in biological media. It also opens the possibility of using intrinsic Raman or parametric gain in other systems, where dissipation hinders the progress of the field and limits applications.

**Introduction**. There is an increasing demand for developing new sensing technologies to detect small molecules, nanoparticles and airborne species[1,2,3,4,5,6,7,8,9,11,12,13]. This interest has been largely driven by the recent developments in nanotechnology, bio-chemical sciences and environmental safety and security. Advancements in nanotechnology have led to the synthesis of massive amounts of nanostructures with different shapes, sizes and materials for novel optoelectronic devices[14], energy harvesting platforms[15] and biomedical diagnostics[16] and therapies[17]. In addition to these manmade nanostructures, usually in the size range of 1-100 nm, there have been massive quantities of nanoparticles released into the environment as end- or by-products of industrial and biological processes. Today both the general public and the scientific community are cautious about their potential risks and toxicological effects[18,19]. For example, nanoparticles with sizes in the range of 1-100 nm can enter cells and may alter the basic cell functions whereas those in the range 1-35 nm can pass the blood-brain barrier and reach the brain; inhaled particles with dry size in the range 20-30 nm may undergo size or shape changes due to change in relative humidity which modifies their deposition patterns in the respiratory system, adversely affecting the human health; and aerosol particles can modify the electromagnetic radiation balance in the atmosphere affecting the cloud properties and formation. Thus, accurate detection, counting and size measurement of nanoparticles and aerosols will contribute to understanding their size dependent toxicity, tailoring their size, designing systems for exposure-prevention, and identifying their benefits and risks on health, technology, and environment. Moreover, developments in biological and chemical sciences have led to the discovery of biomarkers for the diagnosis and prognosis of complex diseases that affect millions of people around the world. Detection of these biomarkers in low concentrations, possibly at single molecule or particle resolution, is crucial for early detection of diseases, developing therapies and monitoring the efficacy of treatment strategies. Finally, homeland security, environmental safety and public health need rapid detection, measurement and identification of threats from airborne contaminants, pandemic viruses, bacteria and bio/chemical warfare agents to design response strategies for handling the risks and searching for appropriate medical treatment in a timely manner. Thus, accurate, rapid and highly sensitive methods for detecting and characterizing nanoscale objects are of urgent need.

The main challenge in sensing technologies is to achieve label-free detection at low concentrations. Microscopy techniques[20,21] have played important roles in attaining super-resolution imaging of nanostructures and even single molecule detection, albeit with the help of optical, chemical or radioactive labels, which may potentially alter the properties and functions of the targeted materials. Label-free techniques, on the other hand, do not use labeling and rely directly on monitoring changes in a sensing signal due to the intrinsic properties of the material such as its refractive index, wavelength dependent absorption, Raman shifts due to its vibrational, rotational or translational states or mass-to-charge ratio. Spectroscopic and imaging techniques have played central roles in single nanoparticle/molecule detection; however, their widespread use is limited by bulky and expensive instrumentation and long processing times.

In the past decade, we have witnessed a boost in the number of label-free detection techniques with varying levels of sensitivities. Techniques relying on the measurement of electrical conductance[22,23,24], light scattering and interferometry[9,13,25,26,27,28,29], surface and localized plasmon resonance[30,31,32], nanomechanical resonators[11,12,33] and optical resonances[2,4,10,34] have been demonstrated. Among these, optical resonators, in particular WGMRs, have become very promising technologies due to their extraordinary sensitivities to changes and perturbations in their structure or proximity[35,36,37,38,39,40,41,42,43,44,45]. They have been successfully employed for sensing biomarkers, DNA, and medium size proteins at low concentrations, as well as for detecting viruses and nanoparticles at single particle resolution. Frequency shift, linewidth broadening or mode splitting has been used as the sensing signal[5, 6,8,46,47].

Detection limit and sensitivity of WGMRs are mainly determined by the strength of light-matter interactions, quantified by the ratio Q/V. Other determining factors are the mechanical fluctuations and changes in thermal environment, the frequency and phase noise of the laser diodes used to scan across resonance modes to estimate the frequency shift, linewidth broadening or mode splitting, and the electrical noise. The effects of the laser noises can be partially eliminated by using passive techniques such as a reference interferometer[42] or by using active stabilization of coupling conditions and frequency locking[44,47], which, however, introduces additional cost and complexity to the sensing system. Improving detection limit by increasing Q/V requires decreasing V or increasing Q. One can increase Q by compensating the losses, and decrease V, for example by fabricating smaller WGMRs. However, one cannot make V infinitesimally small by fabricating smaller WGMRs without decreasing Q, because bending losses increase significantly as the size of the WGMR is decreased below a critical size, eventually leading to decrease of Q. Here two major lines of research have emerged. The first is the hybrid use of WGM and localized plasmons in which highly confined field (smaller V) of the plasmons are utilized to enhance the interaction[35,36,38,41], and ultra-high-Q WGM resonance is used to read-out the changes, which is reflected as a shift in the resonance frequency of the WGM. The second is the Q-enhancement of WGM resonances by compensating losses via optical gain provided by active materials[6,48,49,50], such as rare-earth ions (erbium $Er^{3+}$, ytterbium $Yb^{3+}$,etc) doped into the microresonator. Such resonators are referred to as active resonators. Moreover, by pumping these resonators above lasing threshold, detection limit is further improved beyond what can be achieved by a passive or an active resonator below lasing threshold, because a laser always has a narrower linewidth than the cavity from which the laser is built (i.e., narrower linewidth enables resolving much smaller resonance shift and mode-splitting). However, the former requires preparation and adsorption of plasmonic nanostructures to the WGMR surface while the latter suffers from the facts that most rare-earth ions are not biocompatible and additional processing steps and costs are required.

He *et al.*[6] have shown that nanoparticle binding to a WGM laser splits a lasing mode into two. The amount of splitting can be measured by monitoring the frequency of the heterodyne beatnote generated when the split lasing modes are mixed at a photodetector. Thus a simple yet highly sensitive scheme for detecting nanoparticles was achieved without the need for spectral scanning of a tunable laser diode across the resonance lines to monitor frequency shifts, linewidth broadening or mode splitting. Here, we take an approach with a fundamentally different physical process. Instead of embedding rare-earth-ions as the gain medium in silica microtoroid resonator, we use the intrinsic Raman gain in silica to achieve loss compensation and highly sensitive nanoparticle detection. This does not require any dopant or additional fabrication complexities.

Stimulated Raman scattering is a nonlinear optical process that provides optical gain in a broad variety of materials[51]. Raman process generates photons at a frequency that is up- or down-shifted (anti-Stokes or Stokes photons) (Fig. 1c) from the frequency of the incident photons by an amount equivalent to the frequency of an internal oscillation of the material system, such as vibration, rotation, stretching, or translation. Raman gain has found many applications in biology, material science, sensing, environmental monitoring, optical communication, laser science and spectroscopy[52,53,54,55,56]. However, in most of the material systems Raman gain is very small (of the order of $10^{-13}$ m/W for materials such as silica, silicon, $CaF_2$ etc.), requiring high intensity pump lasers to drive the system above its lasing threshold. A remedy to this challenge is to use field confinement and resonant enhancement of the pump and the emitted photons. Raman lasing has been observed in silicon waveguide cavities[57], silicon waveguides within fiber ring cavities[58], silicon photonic crystal cavities[59], and WGM resonators such as silicon microring[60], silica microspheres[61], silica microtoroids[62,63], glycerol-water droplets[64], and $CaF_2$ disk[65]. However, up to date, Raman gain or Raman lasing has not be used for loss compensation to enhance optical detection capabilities at single particle resolution. Whispering-gallery-mode microtoroidal silica resonators are ideally suited for Raman laser applications because they can be mass-fabricated on silicon chip such that different spectral bands can be covered on a single chip. They have high $Q$ and microscale V which make it possible to easily achieve high intracavity powers to enhance nonlinear effects and obtain low threshold lasing ($P_{threshold} \propto V/Q^2$). Finally they are compatible with optical fibers and can be readily integrated into existing optical fiber networks.

In this paper, we demonstrate Raman-gain induced Q-enhancement (linewidth narrowing via loss-compensation), Raman gain enhanced detection of mode splitting in the transmission spectra, and splitting in Raman lasing for single nanoparticle detection and counting. Our approach of replacing rare-earth-ion doped WGM microlasers by WGM silica Raman microlaser for mode-splitting based nanoparticle detection brings three fundamental improvements. First is the creation of a dopant-free low-threshold microlaser for sensing applications, which retains the inherent biocompatibility of silica. Second is the ability to use the same device as a microlaser with emission in different spectral bands, which can be achieved just by changing the wavelength of the pump laser or by using a broadband pump. Note that in WGM microlasers with rare-earth-ion dopants, one should not only change the dopant but also the pump to obtain emission in different spectral windows. These add fabrication complexities and cost. Third is the record-low size in detecting nanoparticles at single particle resolution using WGMRs. This smallest detected nanoparticles are sodium chloride (NaCl) nanoparticles of radii 10 nm that have smaller polarizabilities than those of polystyrene and gold nanoparticles of the same sizes.

**Results**

**Device fabrication and characterization**

We fabricated fiber-taper coupled silica microtoroid resonators and utilized Raman gain for loss compensation and for WGM Raman microlaser, which were used for ultra-sensitive detection of single-nanoparticle induced mode-splitting. A schematic of the experimental set-up is shown in

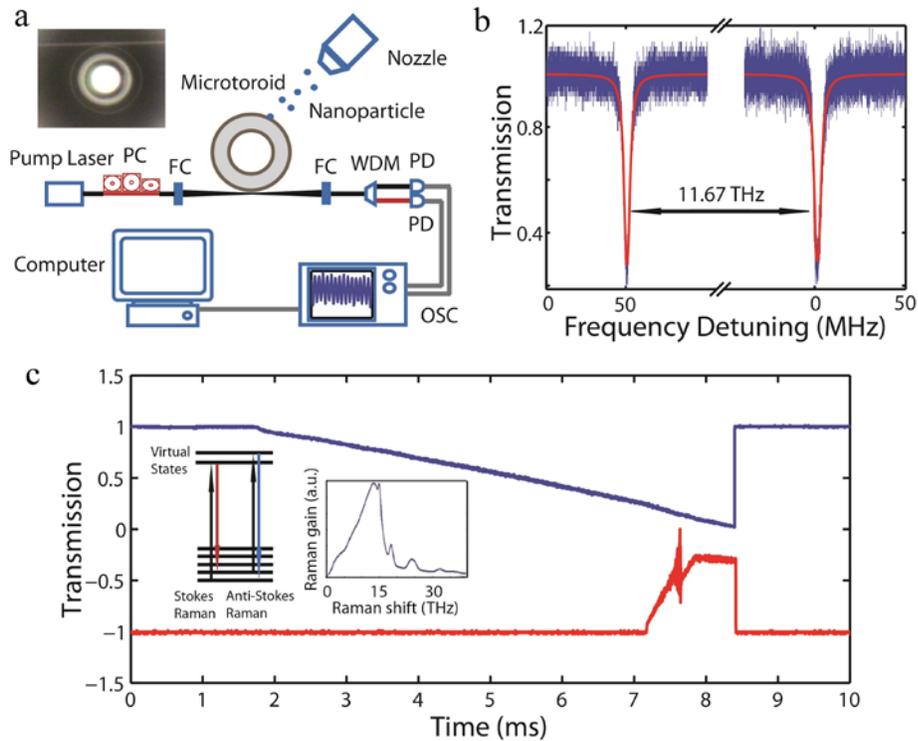

*Figure 1| Experimental setup and measurement method.(a)* Schematic illustration of the experimental setup used in the study of Raman gain enhanced detection of single nanoparticles using mode splitting. A differential mobility analyzer with a nozzle is used to deposit nanoparticles into the mode volume of the resonator one-by-one. Light from a pump laser is coupled to the WGM by a fiber-taper coupler. Polarization of the light was adjusted using a polarization controller (PC). The pump light and the Raman laser light (probe light in case of below lasing threshold operation) are separated from each other using a wavelength division multiplexer (WDM) and detected with photodetectors (PD) connected to an oscilloscope (OSC). The inset shows the top view of a fiber-taper-coupled silica microtoroid resonator taken by an optical microscope. *(b)* Transmission spectra of the silica microtoroid obtained in the 1450 nm and 1550nm bands at low optical power. The resonances in these bands are separated from each other by 11.67 THz which lies within the Raman gain spectra given in the inset of c. *(c)* Typical transmission spectra obtained in the experiments for the pump in 1450 nm band and Raman gain and the laser in the 1550 nm band when the wavelength of the pump laser (blue) is scanned in time. The sawtooth-like waveform is due to the thermal broadening of the resonance line. As we move along this sawtooth-like form in time, more pump light is coupled into the resonator, and cavity power build-up sufficient to produce Raman gain and lasing is achieved as seen in the increased signal in the transmission obtained in the 1550 nm band (red). Insets show the Raman process and the Raman gain spectra for silica.

Fig.1a. Tunable external cavity lasers with emission in the 660nm, 980nm, 1450nm and 1550nm were used as pump lasers to obtain Raman gain and Raman lasing in different spectral bands. The pump light was coupled to microtoroid resonator via a fiber taper and the resultant Stokes photons and laser were out-coupled from the resonator with the same taper. The silicon chip with the silica microtoroids was placed on a 3D nanopositioning stage to precisely tune the distance between the fiber taper and the microtoroid. A fiber polarization controller was used to change the polarization state of the pump laser to maximize the coupling efficiency. The pump lasers were scanned repeatedly through a frequency range of 30 GHz around a single WGM and the transmitted power was detected at the other end of the fiber with photodetectors. Figure 1b shows typical resonance spectra obtained for a silica microtoroid in the 1450 nm pump band and 1550 nm Raman gain band. Typical transmission spectra obtained for the pump (1450 nm) and the Raman signal (1550 nm) as the pump wavelength was scanned are given in Fig. 1c, where we see that as the pump was scanned, cavity power build-up becomes sufficiently high to produce Raman gain and lasing in the 1550 nm band.

**Raman gain enhanced detection of mode splitting in the transmission spectra**

Performance of WGM silica microtoroid resonators as a sensing platform strongly depends on their *Q*-factors which determine not only the field enhancement through Q/V but also the minimum detectable mode-splitting or frequency shift. *Q*-factor is fundamentally limited by the absorption and scattering losses of the light field in the resonator. It has been discussed and experimentally demonstrated that providing optical gain to compensate for a portion of the losses helps to improve *Q*-factor and hence the sensitivity of the resonators[6,48,49]. All previous works along this direction have been performed using rare-earth-ions as optical gain providing dopants[6,48,49,50]. Here, for the first time, we show that Raman gain in silica can be used for the same purpose.

It is known that WGMRs support two counter-propagating modes (clockwise CW and counterclockwise CCW) at the same resonance frequency ω, and that a scattering center (e.g., a nanoparticle, a virus or a molecule) can lift this degeneracy, leading to the splitting of the single resonance mode into two modes, by mediating a scattering-induced coupling between the CW and CCW modes[8,66]. Mode splitting can then be resolved in the transmission spectra of the WGMR if the amount of splitting $2g=-\alpha f^2\omega/V$ is larger than the total loss of the system, quantified by the strict condition $|2g|>\Gamma+\omega/Q$ for well-resolved mode-splitting[67]. Here f is the field distribution of the WGM, $\alpha=4\pi R^3(n_p^2-1)/(n_p^2+2)$ is the polarizability of a particle of radius R and refractive index $n_p$ with the surrounding medium as air, $\Gamma=(8/3)g^2\alpha/\lambda^3$ is the additional loss induced by the scatterer and $\omega/Q$ is the linewidth of the resonance (quantifying loss before the scatterer is introduced). For very small particles we have $\Gamma \ll \omega/Q$ thus the strict condition reduces to $2g> \omega/Q$. In practice, satisfying this strict condition is in general difficult, and the split modes overlap with each other. Although, in principle, we can resolve splittings as small as $\omega/NQ$ where N is a number in the range 10-50 depending on the experimental system and signal processing capabilities, there is a detection limit beyond which the mode-splitting cannot be resolved. The dependence of $2g$ on $f^2$ and $\alpha$ implies that if the overlap between the mode field and the scatterer is not high enough or if the particle is too small, the induced mode-splitting may be so small that it cannot be resolved. In such cases, providing optical gain to increase the Q and hence to reduce the linewidth of the resonance will help to resolve the mode splitting[6,49].

We conducted two sets of experiments to verify the Raman gain assisted Q-enhancement via loss compensation and hence improved detection of mode splitting. In the first experiment, we intentionally induced a small mode-splitting using a fiber tip such that mode splitting could not be resolved by a low-Q resonance in the 1550 nm band. By pumping the silica microtoroid using a laser in the 1450 nm band, we monitored the transmission spectrum in the 1550 nm band. As the pump power was increased, generated Stokes photons compensated for the losses leading to narrowing of the linewidth of the resonance (Fig. 2). As a result initially unobservable mode splitting became clear (Fig. 2a).

In the second experiment, we adjusted the position of the scatterer (fiber tip) in the mode volume such that it introduced a very small amount of mode splitting. Then we set the taper-resonator system to under coupling regime so that the features of the mode splitting is barely seen when the pump laser was turned off. We then turned on the pump laser and increased its power, which revealed a very clear mode splitting of 1.5 MHz in the transmission spectrum due to Q-enhancement (linewidth narrowing). We also observed that the gain shifted the taper-resonator coupling condition from undercoupling to close-to-critical coupling regime. This can be understood as follows. In the undercoupling condition coupling losses quantified by $\kappa_{ext}$ is much smaller than the intrinsic losses $\kappa_o$ (i.e., $\kappa_o \gg \kappa_{ext}$). The induced gain then reduces $\kappa_o$ and brings it closer to $\kappa_{ext}$, and thus shifting the system from undercoupling regime to critical coupling regime where $\kappa_o = \kappa_{ext}$. This is reflected in the transmission spectra as a transition from a close-to-unity transmission to close-to-zero transmission at resonance and better resolved splitting (Fig. 2b).

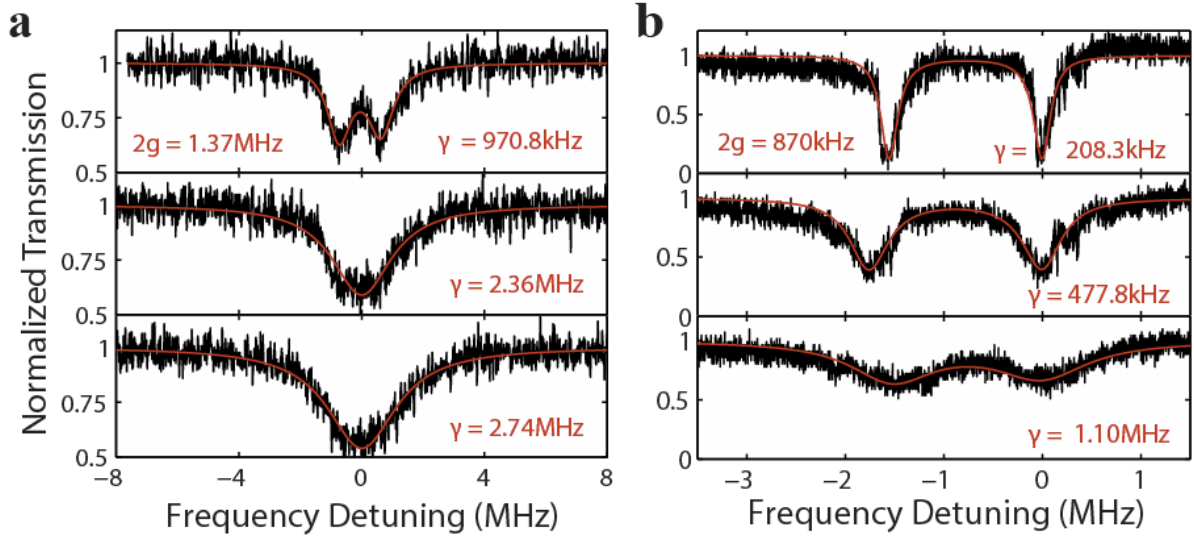

*Figure 2 | Raman gain enhanced detection of scatterer-induced mode splitting. The experiments were performed by a pump-and-probe method. Pump laser in the 1450 nm band was used to provide Raman gain in the probe band of 1550 nm. As the pump power was increased, provided Raman gain increased from bottom to top in **a** and **b** where the spectra in the bottom panels were obtained when the pump was turned off (no gain). (**a**) Initially unseen mode splitting in the transmission spectra (bottom) became visible (top) due to the narrowing of resonance linewidth as the pump power was increased. (**b**) Initially barely seen mode splitting (bottom) became much clearer and well-resolved as the pump power was increased. Moreover, the split resonances become deeper, implying that the taper-resonator systems moves from undercoupling regime to closer to critical coupling.*

**Raman lasers and scatterer-induced splitting of lasing lines**

Once the pump power exceeds a threshold value, lasing modes could be observed at frequencies red-shifted relative to the pump frequency (i.e., approximately 12THz and 14 THz for pump wavelengths in the 1450 nm and 1550 nm, respectively). At much higher pump power, we observed that the spectrum evolved from a single mode Raman laser into a spectrum showing multiple Raman lasing peaks separated by the free spectral range of the microtoroid, as well as cascaded Raman lasing separated by Raman shift. We obtained lasing in different spectral bands using the same microtoroid but different pump lasers, implying that a single WGM microresonator can be used to generate lasing at different colors covering a large range of spectrum and hence can be used for optical detection and sensing in all these bands. It is important to note here that in rare-earth-ion-doped resonators one should prepare resonators with different dopants for different spectral bands of interest. For example, erbium- or ytterbium-doped resonators are used for lasing in 1550 nm or 1040 nm bands, and thulium- or neodymium-doped resonators are used for lasing in the visible band. However, using the Raman process one can obtain lasing in different spectral bands using the same resonator (Fig. 3). Moreover, by fine-tuning pump power and/or the coupling strength between the fiber-taper and the resonator one can get single or multimode lasing (Fig. 3) covering a large band of the spectrum and wavelength regions that are not accessible with conventional lasers.

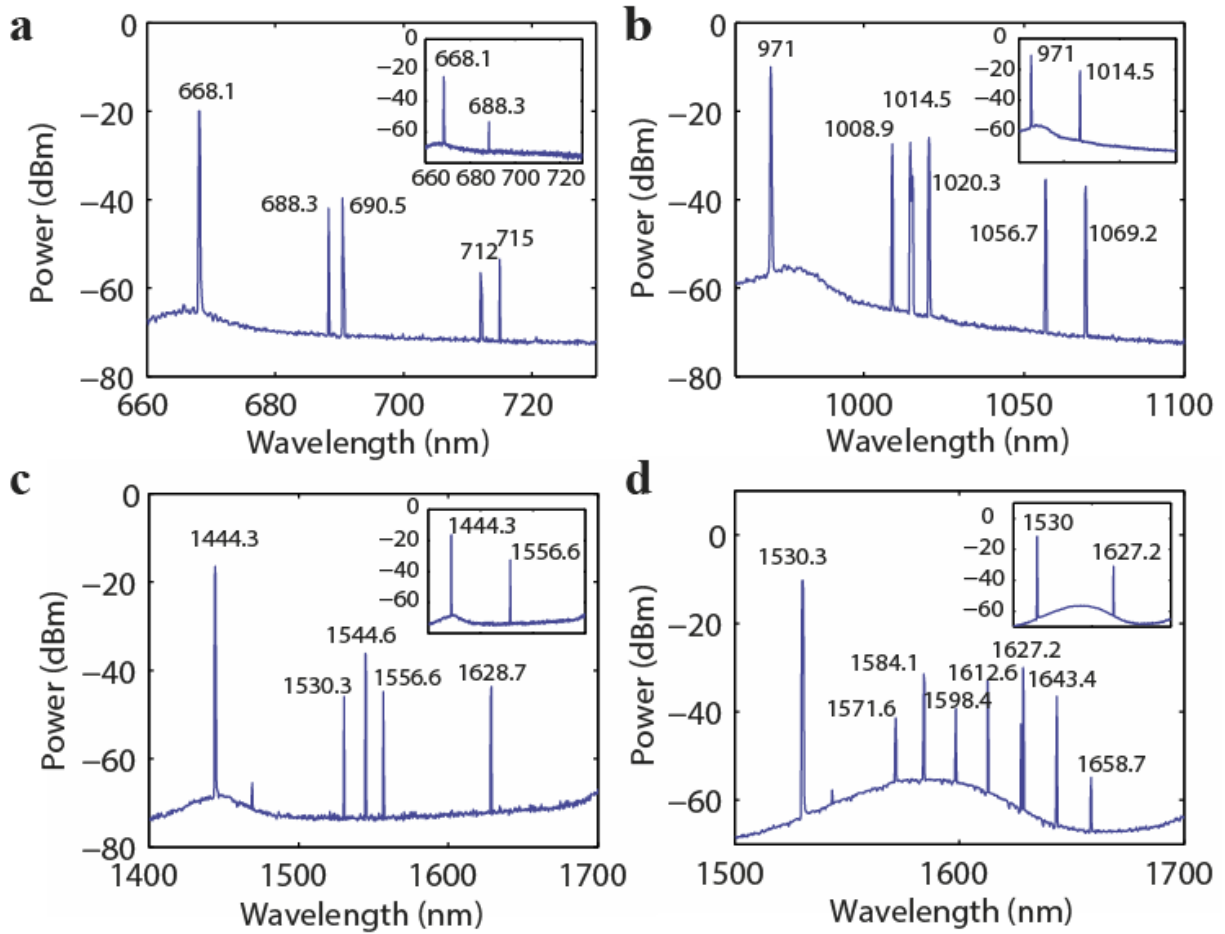

*Figure 3 |Emission spectra of Raman lasing obtained in the same silica microtoroid resonator at different bands of the spectrum covering visible to near-infrared.* The pump lasers are in (**a**) 660 nm,(**b**) 980 nm, (**c**) 1450 nm, and (**d**) 1550 nm bands. Higher order cascaded Raman Stokes lasing is clearly seen in all the spectra. Insets show the single mode operation of the Raman lasers obtained by tuning the pump power and the coupling condition.

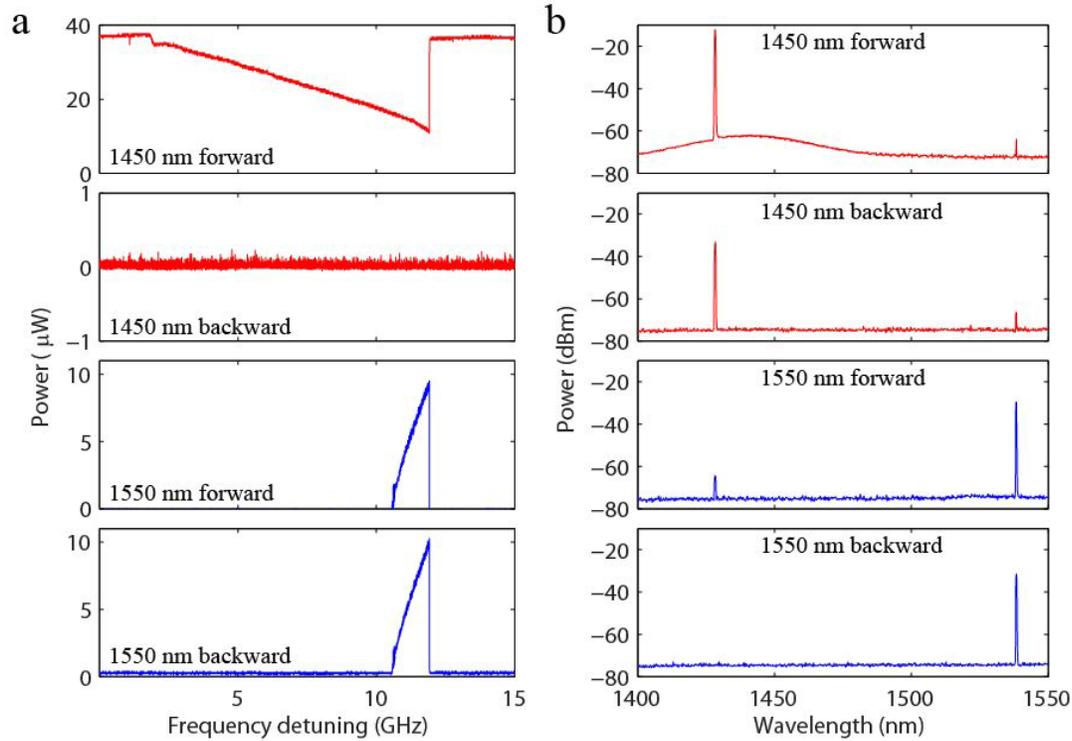

*Figure 4| Experimental results showing the bidirectionality of Raman emission.(a)Transmission spectra and (b) optical spectrum obtained in the forward and backward directions for the pump laser in 1450 nm and the Raman emission in the 1550 nm band as the wavelength of the pump laser was scanned. The peaks observed in b for the backward direction of 1450 nm band is due to the scattering of 1450 nm pump light from the connectors, circulators and WDM used in the circuit, and these peaks appear even when the resonator is removed from the setup. This is 20dB smaller than that in the forward direction. The small peak in the 1450nm band observed in the 1550 nm forward direction is due to the crosstalk between the channels at the WDM. Absence of a beatnote signal in the 1550 nm backward and forward in a implies that there is no observable mode splitting and therefore Raman emission is bidirectional. Absence of a resonance peak in the backward direction for the 1450 nm pump implies that there is no observable mode splitting in the 1450 nm band and pump is unidirectional.*

Starting with microtoroid resonators without observable mode splitting, we observed that Raman effect excited two counter-propagating whispering-gallery laser modes, i.e., CW and CCW modes, both of which are eigenmodes of the cavity (Fig. 4). Similar to the mode splitting process observed in passive and active resonators excited with pump powers below lasing threshold, a scatterer within the mode volume can couple these counter-propagating lasing modes to each other leading to the scatterer-induced splitting of Raman laser. When these split lasing modes are mixed at a photodetector of sufficient bandwidth, a heterodyne beat note signal with frequency corresponding to the amount of splitting is obtained. We confirmed the bi-directionality of Raman emission by performing an experiment using 1450 nm pump laser to excite Raman lasing in the 1550 nm band. We input pump light into the resonator in the CW direction (forward) and monitored the transmission spectra in both the forward and backward (CCW) directions. Absence of a resonance peak in the backward direction for the 1450 nm pump implied that pump stayed in the forward direction (Fig. 4a) and there was no scattering center that could lead to mode splitting in this band and channel a portion of the light into the backward direction. With this unidirectional input, we monitored the backward and forward directions for the 1550 nm band and observed optical power in both directions, implying that the Raman emission existed in both directions (bidirectional). Since there was no beatnote oscillation in the 1550 nm band (Fig. 4a), we concluded that there was no observable mode splitting, and that the Raman emission naturally couples to both directions in the resonator. We also measured the optical spectrum in the forward and backward directions (Fig. 4b) confirming the results observed in the transmission spectra.

Next we confirmed the generation of a beatnote signal due to scatterer-induced mode splitting by introducing a nanofiber tip as the scatterer into the mode volume of the microlaser and by monitoring the heterodyne beatnote signal in response to the positions of the scatterer in the mode volume (Fig. 5). We performed the same experiments using lasing lines at different wavelength bands but in the same resonator. We observed that the beatnote signal and its frequency are not only affected by the size of the scatterer but also by the overlap between the scatterer and the field of the lasing lines. As seen in Fig. 5, at a fixed location of the scattering center, the amount of splitting experienced by Raman lasers at different spectral bands is different from each other. Moreover, each scatterer may increase or decrease the splitting experienced by each lasing line depending on the distribution of the scatterers within the mode volume of each lasing mode. Thus, with each new scatterer entering the resonator mode volume, splitting may increase or decrease for all lasing modes in different bands or may decrease for some lasing modes while increase for the others. The amount of change in the splitting is different for different lasing modes. These can be attributed to the facts that spatial overlaps between the scatterers and the fields of different lasing modes are different, and that mode splitting scales inversely with the mode wavelength. This observation demonstrates that making measurements at multiple wavelengths or spectral bands enables us to detect nanoparticles or scattering centers that could have gone undetected if only a single lasing mode were used.

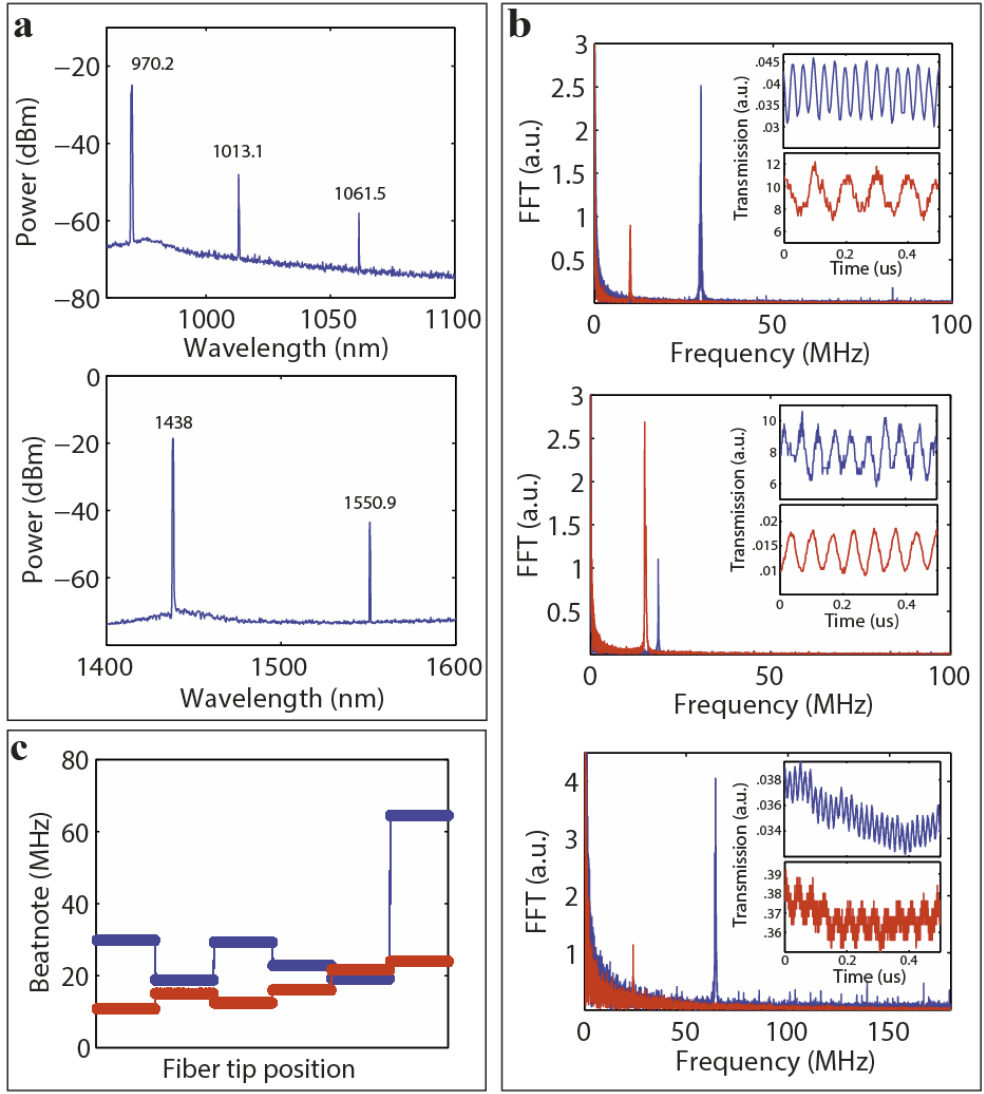

*Figure 5| Detection of scatterer-induced mode splitting using WGM Raman lasing in two different bands in the same silica microtoroid.* The color codes blue and red correspond to the experiments performed with pump laser in the 980 nm (blue) and 1450 nm (red) bands. **(a)** Optical spectra of Raman lasers with pumps at 980 nm (upper) and 1450 nm (lower) bands. **(b)** Change in the beatnote signal (inset) and its frequency obtained using Fast Fourier Transform (FFT) when the fiber tip position or size changes. **(c)** Changes in the beatnote frequency as the fiber tip repeatedly enters and exits the mode volume, each time at a different position with different effective tip size in the mode volume. The response of the lasing modes at different bands are different. Splitting of the lasing modes (beatnote frequency) may increase or decrease depending on how the scatterers are distributed within the mode volume of each mode: Splitting may increase or decrease for both lasers in different bands or may decrease for one laser while it increases for the other laser.

**Demonstration of single nanoparticle detection**

We evaluated the performance of WGM Raman microlaser and the mode splitting method to detect nanoparticles with single particle resolution. In these experiments, we used a differential mobility analyser (DMA) accompanied by a nozzle to deposit nanoparticles onto the microlaser. Nanoparticles were carried out from their colloidal solution using a Collison atomizer[8]. After the evaporation of the solvent in polydisperse droplets, the solid particles were neutralized to maintain a well-defined charge distribution. Then they were sent to the DMA which classified them according to their electrical mobility. The output slit of the DMA allows only the particles within a narrow range of sizes to exit and land on the WGM microlaser via the nozzle. The flow rate and the concentration of the colloidal solution were set low to ensure deposition of particles one-by-one onto the microlaser.

We tested our system using gold (Au), polystyrene (PS) and sodium chloride (NaCl) nanoparticles. As discussed in the previous section, particle binding to the WGM microlaser led to the splitting of a lasing line into two which eventually gave a heterodyne beatnote signal when mixed at a photodetector. The beatnote frequency corresponds to the amount of splitting. Each consecutive nanoparticle binding event led to a discrete change in the beatnote frequency. The frequency may increase or decrease depending on the location of each particle with respect to the field distribution of the lasing modes and the position of the particle with respect to previously deposited particles in the mode volume[6,8,42,43,49,66]. In Fig. 6, we give the change in beat frequency and hence the splitting of the lasing mode as NaCl nanoparticles of size R=15 nm (Fig. 6a), 20nm (Fig. 6c) and 25 nm (Fig. 6e) were continuously deposited onto the WGM Raman laser. With each particle binding event, we observe a discrete up or down jump in the beat frequency. From the histograms shown in Figs. 6b, 6d and 6f, we see that the larger the particles are, the wider the distribution of the measured changes in the beatnote frequency.

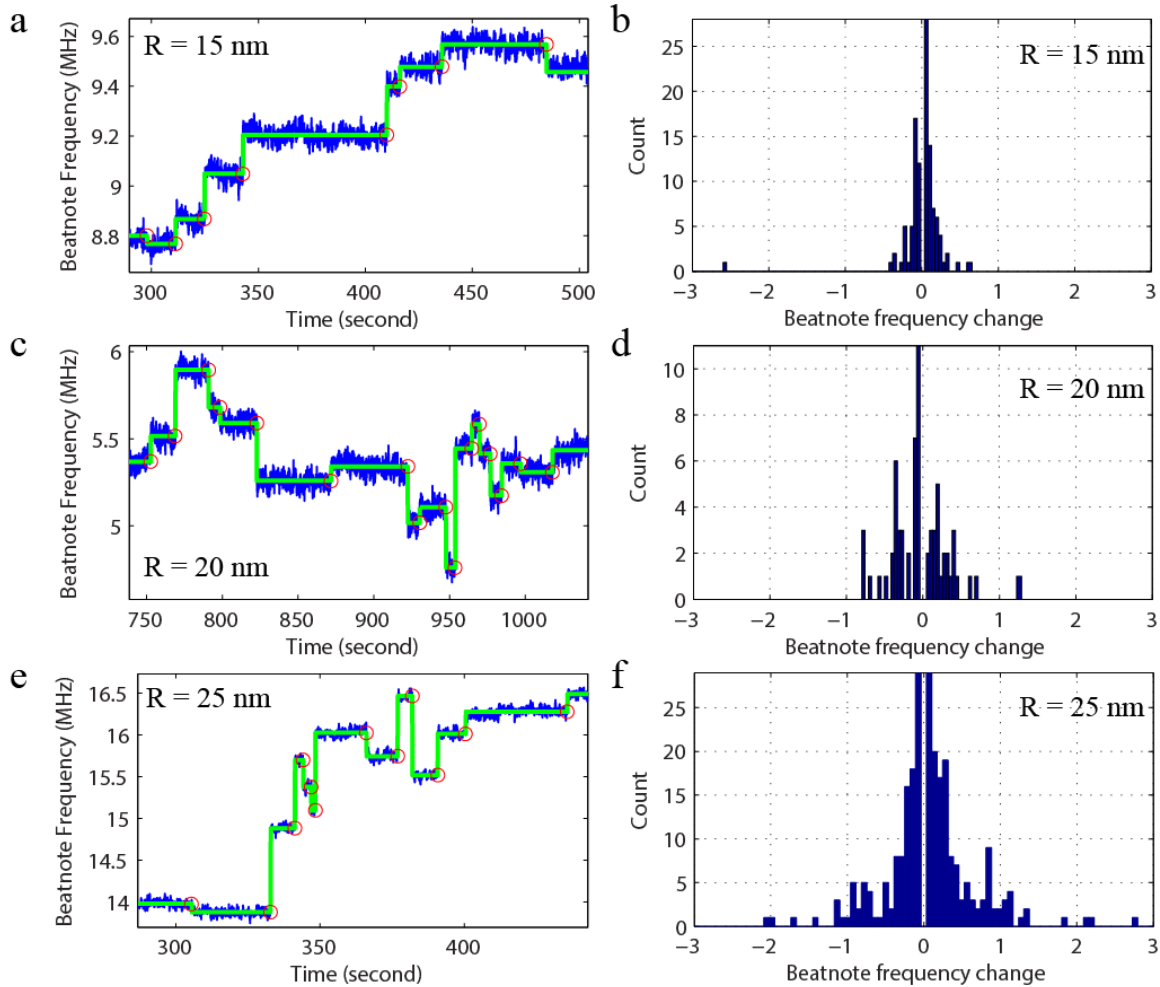

***Figure 6 | Detection of NaCl nanoparticles using the scatterer-induced splitting of WGM Raman laser.** Each discrete upward or downward jump in the beatnote frequency spectra corresponds to a binding event of one nanoparticle with radius **(a)** R=15 nm, **(c)** R=20 nm and **(e)** R=25 nm corresponds to one nanoparticle binding event. Histograms of the beat frequency changes for each nanoparticle binding event for NaCl nanoparticles of size **(b)** R=15 nm, **(d)** R=20 nm and **(f)** R=25 nm imply a correlation between the size of the particle and the width (standard deviation or root-mean-square) of the distribution.*

In order to estimate the reproducibility of the measured beatnote frequency, we stopped nanoparticle deposition at some point and continuously measured beat frequency for extended durations of time. Figures 7a and 7b depicts the beat frequency as a function of time and the histogram of measured frequencies, respectively. The measured frequencies stayed within ±50 kHz of the mean frequency. We also calculated the Allan deviation, which is a commonly used technique to estimate the frequency stability, for beat frequency. We calculated the Allan deviation from segments (integration time) from 0.1 to 200 s. The result is shown in Fig. 7c. In Fig. 7d we present the measurement results for NaCl nanoparticles of R=10 nm. It is seen that while some of the particle binding events led to resolvable changes in the beat frequency, the changes in some other were not very clear and obstructed by the noise level in our system.

Considering that we can resolve some of the binding events even at the present noise level without any active or passive stabilization procedure, we can safely say that the detection limit of the WGM Raman microlaser is 10 nm, which can be further improved by reducing the noise level in the system.

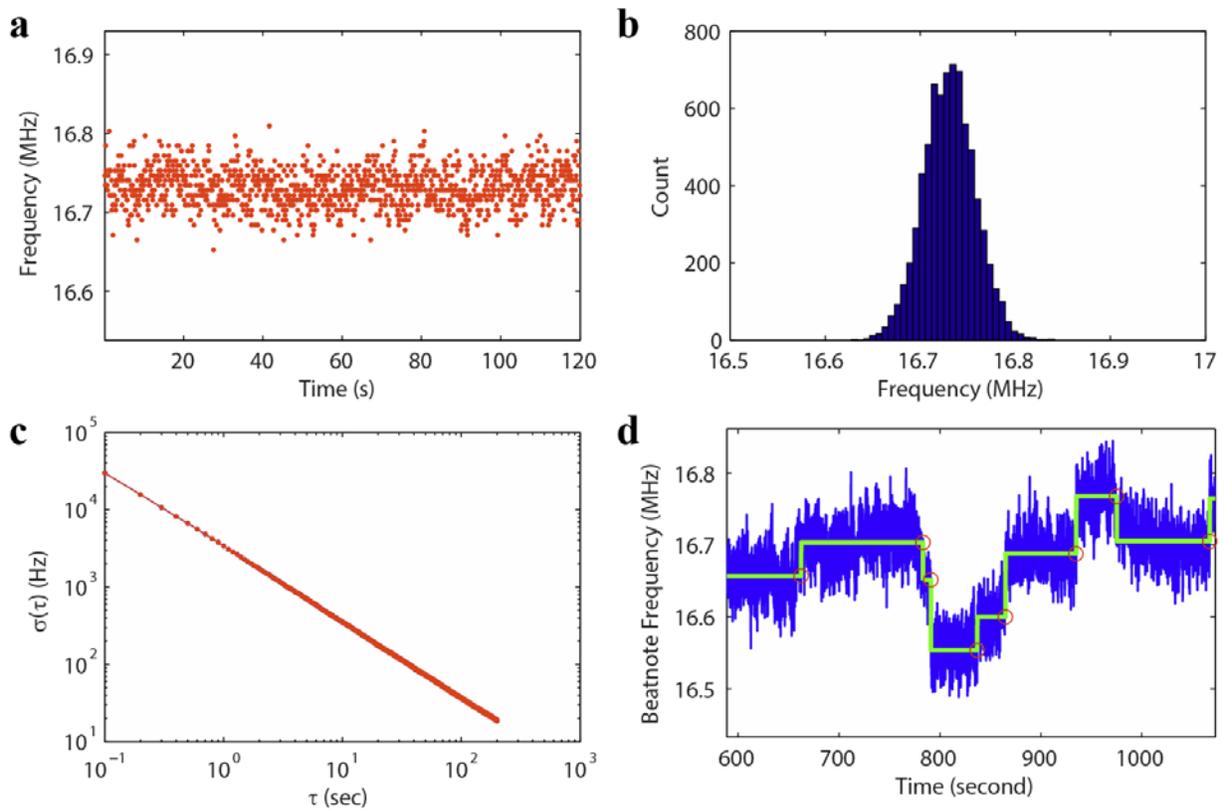

*Figure 7 | Noise analysis. (a) Change in the measured beatnote frequency as a function of time. (b) Histogram of the measured beatnote frequency. (c) Allan deviation of measured frequency as a function of time. (d) Beatnote frequency measured for NaCl nanoparticles of size R=10 nm.*

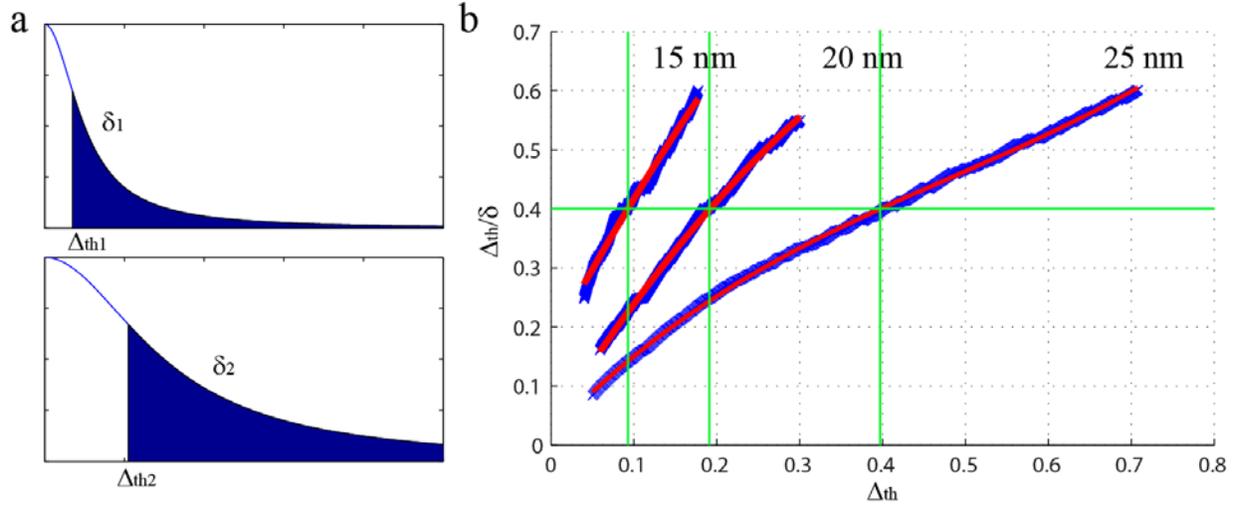

*Figure 8 | Measurement of an ensemble of nanoparticles using scatterer induced beatnote frequency changes of a WGM Raman microlaser.(a)* The distribution of the discrete changes in Raman laser beatnote frequency. Larger particles induce larger changes with wider distribution. However, the shapes of beatnote frequency-change distributions are the same. *(b)* The relations between $\Delta_{th}/\delta$ and $\Delta_{th}$ for detected NaCl particles of radii 15, 20 and 25 nm (from left to right). At the same $\Delta_{th}/\delta$ ratio, the corresponding $\Delta_{th}$ value gives the estimated width of distribution of beatnote frequency changes.

**Size estimation**

As shown in Fig. 6, each discrete change of beat note frequency signals a nanoparticle entering the mode volume. The heights of the discrete frequency jumps are related to the polarizability of the particle and particle positions in the field[6,8]. Different heights for the same particle size are results of different overlaps between the WGM and the detected particles. The distribution of discrete jumps contains information of particle sizes[6]. As seen in Fig. 6b,d,f, larger particles give a wider distribution for beatnote frequency jumps. Since we used the same particle delivery system, the distributions of particle positions in the resonator mode volume are the same for different particle sizes, and the shapes of distribution of the beatnote frequency jumps are the same for different particle sizes. However they cannot be directly compared by calculating the standard deviation of the detected jumps because the jump events below noise level cannot be obtained. To extract the size information, we calculated the root mean square (RMS denoted as δ here) of the beatnote frequency changes that are above a threshold value $\Delta_{th}$. For different particle sizes, the distributions of the changes follow the same statistical model, the ratio of $\Delta_{th}/\delta$ should be equal when $\Delta_{th}$ is at the same position with respect to each distribution (Fig. 8a). Therefore by plotting the $\Delta_{th}/\delta$ for different $\Delta_{th}$ values, one can estimate the effective width of the respective jump distributions. Figure 8b shows the curves of $\Delta_{th}/\delta$ vs $\Delta_{th}$ for three different sizes of NaCl particles. By comparing the scaling of horizontal axis for all three cases, we can extract the ratio of the distribution widths. This suggests that by using measurement results from an ensemble of particles with known sizes one can employ a referencing scheme to assign an average size to a given ensemble of particles: For the reference particle ensemble and the particle ensemble of interest, one can first obtain $\Delta_{th1}/\delta_1$ and $\Delta_{th2}/\delta_2$, respectively, as a function of $\delta_1$ and $\delta_2$ from the measured data, and then find $\Delta_{th1}$ and $\Delta_{th2}$ which satisfy $\Delta_{th1}/\delta_1=\Delta_{th2}/\delta_1$, (In Fig. 8, we use $\Delta_{th1}/\delta_1=\Delta_{th2}/\delta_2=0.4$). Since the discrete jump heights are related to particle polarizability which is proportional to the $R^3$ where R is the radius of detected particles, we can accordingly estimate the size ratio of the three cases. Thus the size of the ensemble can be estimated from $\Delta_{th1}/\Delta_{th2}=(R_1/R_2)^3$. Using this method, we estimate the size ratio to be 30.6:39.3: 50.0 which represents less than 3 percent error compared to nominal values.

**Discussion**

Many different optical techniques and schemes have been proposed and demonstrated over the past decade for detecting, counting and characterizing nanoscale objects with the ultimate aim of reaching single molecule resolution. Whispering-gallery-mode resonators, photonic crystal microcavities and surface or localized plasmons have been the forerunners in these efforts. However, these resonance-based techniques suffer from losses caused by the absorption in the material system from which the sensors are fabricated, and that in the surrounding medium the field is probing. Minimizing the losses due to surrounding requires smaller mode volume V (tighter field confinement) which also minimizes the probing area for detection and sensing. As we have discussed within the realm of WGM resonators, decreasing mode volume below a critical volume comes with a reduction in the Q-factor, implying additional losses in the form of radiation or bending losses. Larger probing area in the surrounding is usually associated with larger mode volume and additional losses due to the surrounding. These unavoidably leads to lower Q/V which decreases the strength of light-matter interactions and hinders achieving the single particle or single molecule detection.

Over the years, several proposals to compensate losses in these photonic devices have been made. A common point in all these proposals, regardless of whether the system is WGM, photonic crystal or plasmonic, has been the incorporation of an active (gain) media into these systems. Among the preferred media are dyes, rare-earth-ions and quantum dots as dopants into the device structure. In this work, we have demonstrated that intrinsic gain such as Raman gain in the materials forming the device structure (WGM resonator, plasmonics, photonic crystal, metamaterial) can be used effectively to compensate the losses in the photonic devices to improve their performances by eliminating the drawbacks associated with the losses and dopant-incorporation into these devices.

In conclusion, we demonstrated that Raman gain in silica WGM resonators can be used to compensate losses to enhance quality factor, and to enable gain-enhanced detection and characterization of nanoparticles at single nanoparticle resolution using the mode-splitting method. We also provided results confirming the bidirectionality of the Raman laser in the WGM resonator. The detected 10 nm NaCl nanoparticle is the smallest nanoparticle detected with optical techniques without labeling. Our experiments presented here have been performed in dry environment. However, recent demonstrations of particle induced mode-splitting[45,68,69] and WGM Raman lasing[70] in liquid environment imply that the techniques developed here can be extended to loss compensation of these devices in liquid environment and biosensing in biological fluids. Moreover, similar to what we have demonstrated here for a silica microtoroid (Raman gain in silica for loss compensation and for improving the detection limit of WGM resonators), Raman gain in materials, which are used to fabricate photonic crystals, plasmonic and metamaterial structures, and as well as other types of WGMRs, can also be used to compensate for their losses and enhance their performance. For example, Raman gain in silicon can be used for loss compensation in silicon microrings and silicon photonic crystal structures. The ideas developed here can be extended to parametric gain in silica and other materials for loss compensation. We believe that this dopant-free loss compensation technique will find applications in other photonic devices and can be effectively used to improve the performance of photonic devices and enhance the detection limits of sensors based on resonance effects. Achieving the detection of nanoparticles down to 10 nm in size and counting them one-by-one are encouraging for us to continue the research to push the detection limit to single molecule resolution using the proposed loss-compensation technique.

**Acknowledgement**

This work was supported by the NSF under grant number 0954941 and the US Army Research Office under grant number W911NF-12-1-0026.

**Author contributions**

S.K.O and J.Z contributed equally to this work. S.K.O and L.Y conceived the idea and designed the experiments, J.Z and X.Y performed the experiments with help from L.H, B.P, H.Y, S.K.O and F.M. Data was analyzed by S.K.O, J.Z, G.L.L and L.Y, and S.K.O and L.Y wrote the manuscript with contributions from all authors.